\documentclass[aps,floatfix,footinbib,amsmath,amssymb,pra,twocolumn,superscriptaddress,longbibliography]{revtex4-2}

\usepackage{xcolor, amssymb, graphicx, amsfonts,epstopdf,float}
\usepackage{multirow,amssymb,amsbsy,amsmath,epstopdf}
\usepackage{hyperref}
\usepackage{verbatim}
\usepackage{bm}
\usepackage{amsthm}
\usepackage{graphicx}  

\makeatletter

\def\printauthornotes#1{%
  \protected@edef\@tempa{#1}%
  \expandafter\printauthornotes@i\@tempa;\@nil
}

\def\printauthornotes@i#1;#2\@nil{%
  \frontmatter@footnote{#1}%
  \def\@tempa{#2}%
  \ifx\@tempa\@empty\else
    \textsuperscript{,\,}%
    \expandafter\printauthornotes@i#2\@nil
  \fi
}

\def\doauthor#1#2#3{%
  \ignorespaces#1\unskip\@listcomma
  \begingroup
   #3%
  \@if@empty{#2}
    {\endgroup{}{}}
    {\endgroup{\comma@space}{}\printauthornotes{#2}}%
  \space \@listand
}
\makeatother

\newcommand{\ket}[1]{\ensuremath{\left|#1\right\rangle}}

\newcommand{\ketbra}[2]{\ensuremath{\left|#1\rangle\langle#2\right|}}

\newcounter{suppnote}
\newcommand{\suppnote}[1]{%
  \refstepcounter{suppnote}%
  \section*{\NoCaseChange{Supplementary Note \arabic{suppnote}: #1}}%
}

\begin{document}

\title{Observing mixed-state Uhlmann winding numbers in integrated photonic quantum walks}

\author{Baichu Yu}
\thanks{These authors contributed equally.}
\affiliation{Quantum Science Center of Guangdong-Hong Kong-Macao Greater Bay Area, Shenzhen 518045, China}

\author{Qin-Qin Wang}
\thanks{These authors contributed equally.}
\affiliation{School of Physics, Hefei University of Technology, Hefei 230009, China}

\author{Huilong Huang}
\thanks{These authors contributed equally.}
\affiliation{Quantum Science Center of Guangdong-Hong Kong-Macao Greater Bay Area, Shenzhen 518045, China}
\affiliation{Shenzhen University, Shenzhen 518060, China}

\author{Bin-Tong Yin}

\author{Xiao-Ye Xu}
\affiliation{Laboratory of Quantum Information, University of Science and Technology of China, Hefei 230026, China}
\affiliation{Anhui Province Key Laboratory of Quantum Network, University of Science and Technology of China, Hefei 230026, China}

\author{Feng Gao}
\email{gaofengsummit@gmail.com}
\affiliation{Peng Cheng Laboratory, Shenzhen, 518000, China}

\author{Xin Wang}
\affiliation{Department of Physics, City University of Hong Kong, 83 Tat Chee Avenue, Kowloon, Hong Kong SAR, China}
\affiliation{City University of Hong Kong Shenzhen Research Institute, Shenzhen, Guangdong 518057, China}

\author{Keren Li}
\affiliation{Shenzhen University, Shenzhen 518060, China}

\author{Sen Yang}
\affiliation{Department of Physics, The Hong Kong University of Science and Technology, Hong Kong, China}

\author{Yuan-Yuan Zhao}
\email{zhaoyuanyuan1@quantumsc.cn}
\affiliation{Quantum Science Center of Guangdong-Hong Kong-Macao Greater Bay Area, Shenzhen 518045, China}

\begin{abstract}
Integer winding numbers provide a well-established classification of topological phases protected by chiral symmetry at zero temperature, but extending this classification to thermal mixed states remains challenging.
The Uhlmann phase, a mixed-state extension of the Berry phase, provides a geometric framework for characterizing finite-temperature topology. 
However, this geometric phase is restricted to two quantized values and therefore cannot resolve the multiple thermal regimes associated with high winding numbers.
Here, we experimentally observe integer mixed-state winding numbers in a photonic quantum-walk chip.
We prepare and reconstruct thermal states associated with high-winding chiral Hamiltonians and their spectrum-preserving dimensional extensions. 
The reconstructed Uhlmann overlaps form closed trajectories in the complex plane, and their integer windings around the zero-overlap point are directly observed.
As temperature increases, the Uhlmann winding number undergoes multiple discrete changes, accompanied by switches of the Uhlmann phase and the successive disappearance of momentum-resolved topological kinks.
These results highlight an integer-valued topological characterization alongside the mixed-state geometric phase.
\end{abstract}

\maketitle

\noindent
{\bf\large Introduction}

\noindent
Topological phases are characterized by global geometric structures encoded in Bloch states\,\cite{Schnyder2008,Chiu2016}.
For chiral-symmetric Bloch Hamiltonians, this structure is encoded along a closed momentum loop by an integer winding number that counts how many times the unit Bloch vector winds around the origin within the chiral plane. 
Because the associated Zak--Berry phase\,\cite{Atala2013zak,Cardano2017,Wang2023berrymobius} is defined modulo $2\pi$, it retains only the parity of the winding number, whereas the integer invariant resolves the full high-winding structure\,\cite{Berry1984,Zak1989}. 
This integer classification underlies the Su--Schrieffer--Heeger model\,\cite{Su1979} and its long-range generalizations\,\cite{Beatriz2019}, in which high winding numbers have been engineered\,\cite{Xu2018,Xiao2018}. 
At finite temperature, however, equilibrium states are represented by density matrices rather than occupied-band wavefunctions, requiring a geometric framework formulated directly for mixed states. 
How the full integer structure associated with high winding evolves with temperature remains unclear.

Geometric phases have been extended to mixed states and open-system dynamics through several formulations\,\cite{Uhlmann1986,Sjoqvist2000,Tong2004,bardyn2018probing,Unanyan2020,Cohen2019}.
The Uhlmann construction extends the holonomy structure underlying the Berry phase by introducing parallel transport for density-matrix amplitudes\,\cite{Uhlmann1986}.
For a closed momentum loop, the phase of the overlap between the initial amplitude and its parallel-transported counterpart defines the Uhlmann phase, which has served as a finite-temperature topological indicator in a range of one- and two-dimensional models\,\cite{viyuela2014uhlmann,Huang2014,viyuela2014two,Viyuela2015,Bruno2017}.

In a representative class of chiral two-band systems, the Uhlmann overlap is real, so the Uhlmann phase takes one of two quantized values\,\cite{viyuela2014uhlmann}. 
As temperature varies, zeros of the overlap mark transitions between these values.
High-winding systems can consequently exhibit multiple temperature-driven Uhlmann-phase transitions\,\cite{hou2021finite,morachis2021topological,zhang2021comparison}. The corresponding parallel transport can develop multiple singular phase reversals along momentum, known as topological kinks\,\cite{viyuela2014uhlmann}.
Because the Uhlmann phase records only the parity of these reversals, distinct intermediate thermal regimes can share the same value.

Uhlmann-based integer indices have instead been constructed from the spectral flow of complex holonomy eigenvalues and from the winding of the Uhlmann phase across a second crystalline momentum\,\cite{Huang2014,viyuela2014two,Viyuela2015}.
Existing experiments have measured the mixed-state geometric phase in nuclear-magnetic-resonance systems\,\cite{Du2003,Zhu2011} and observed topological Uhlmann-phase transitions using superconducting-qubit\,\cite{viyuela2018observation,mastandrea2026intermediate}, bulk-optical\,\cite{wang2026mixed} and electrical-circuit simulators\,\cite{Huang2026Electrical}.
However, the integer winding structure of Uhlmann geometry and the multiple thermal regimes it distinguishes have not been directly observed experimentally.

\begin{figure*}[t]
    \centering
    \includegraphics[width=0.95\linewidth]{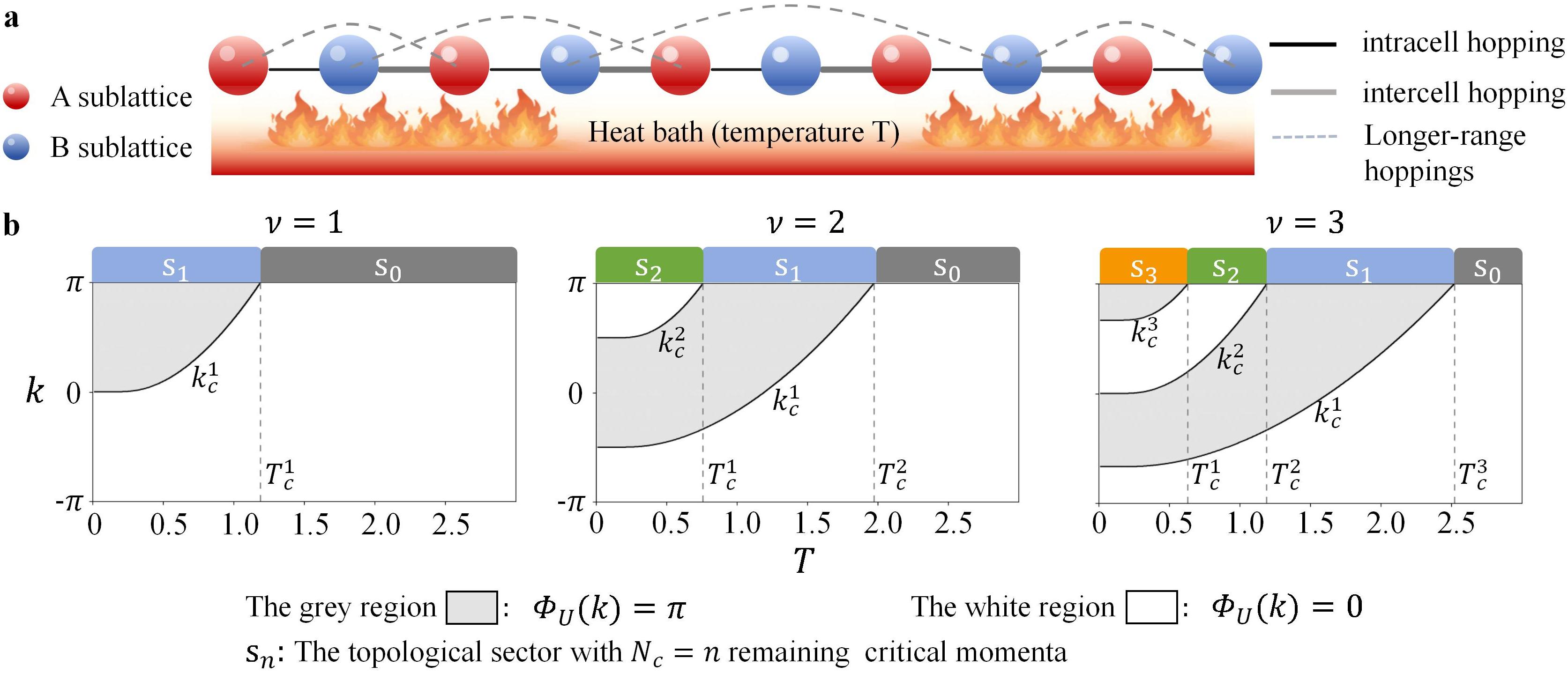}
    \caption{\textbf{Schematic illustration of multiple thermal topological transitions.}
    \textbf{a}, A 1D chiral-symmetric generalized SSH chain coupled to a heat bath at temperature $T$. The red and blue sites denote the $A$ and $B$ sublattices, respectively. The dashed curves represent long-range intersublattice hopping, which enables higher winding numbers while preserving chiral symmetry.
    \textbf{b}, Thermal evolution of representative flat-band Hamiltonians with chiral winding numbers $\nu=1,2,3$. The black curves trace the critical momenta $k_c^{(j)}(T)$ at which the accumulated Uhlmann phase $\Phi_U(k,T)$ changes discontinuously by $\pi$, forming topological kinks. As the temperature increases, these kinks move toward $k=\pi$ and disappear successively at the critical temperatures $T_c^{(j)}$. Accordingly, the Uhlmann winding number $\nu_U$ decreases successively to zero, while the full-loop Uhlmann phase $\Phi_U(T)$ alternates between $0$ and $\pi$. The upper colored bars denote the sectors $S_n$ with $n$ remaining critical momenta and $\nu_U=n$. The gray and white regions correspond to $\Phi_U(k,T)=\pi$ and $0$, respectively.}
    \label{fig.system}
\end{figure*}

Here, we experimentally access the integer winding structure of finite-temperature Uhlmann geometry using a programmable integrated photonic chip.
We characterize this geometry through two complementary representations of Uhlmann parallel transport.
One resolves the accumulated transport along the physical momentum loop, revealing momentum-resolved topological kinks.
The other tracks the full-loop Uhlmann overlap across a spectrum-preserving dimensional extension, yielding a closed trajectory in the complex plane whose winding defines the Uhlmann winding number.
We implement both characterizations on the quantum-walk chip by preparing thermal states of the chiral Hamiltonians and their dimensional extensions, then reconstructing the corresponding momentum-resolved density matrices.
As temperature increases, we observe that each transition removes one kink from the physical momentum loop and reduces the winding number of the extended-overlap curve by one, whereas the Uhlmann phase records only the parity of this integer.

~\\
\noindent
{\bf\large Results}

\noindent
\textbf{Berry phase and chiral winding number}

\noindent
Here, we consider a family of one-dimensional (1D) two-band topological systems [Fig.\,\ref{fig.system}(a)] described by generalized Su-Schrieffer-Heeger (SSH) Hamiltonians with long-range intersublattice hopping\,\cite{chen2020elementary}.
Their momentum-space Hamiltonian can be written as:
\begin{equation}\label{Hamiltonian}
H_k
=
\mathbf d_k\cdot\boldsymbol{\sigma}
=
E_k\hat{\mathbf n}_k\cdot\boldsymbol{\sigma},
\end{equation}
where the two bands have energies $\pm E_k$, $\hat{\mathbf n}_k$ is a unit Bloch vector, and $\boldsymbol{\sigma}=(\sigma_x,\sigma_y,\sigma_z)$ is the vector of Pauli matrices. 
At zero temperature, the occupied band $\{\ket{u_{g,k}}\}$ carries the Zak--Berry phase:
\begin{equation}\label{BerryPhase}
\Phi_B
=
i\oint_{\mathrm{BZ}}
\langle u_{g,k}|\partial_k u_{g,k}\rangle\,dk.
\end{equation}

The models considered here possess chiral symmetry that confines $\hat{\mathbf n}_k$ to a fixed great circle of the Bloch sphere.
Choosing an oriented orthonormal basis $(\mathbf e^i,\mathbf e^j)$ in this plane, we write $\hat n_k^i+i\hat n_k^j=e^{i\alpha_k}$, where $\alpha_k$ is continuously unwrapped along the Brillouin zone. 
The resulting chiral winding number is:
\begin{equation}\label{WindNum}
\nu
=
\frac{1}{2\pi}
\oint_{\mathrm{BZ}}d\alpha_k
\in\mathbb Z,
\quad
\Phi_B=\pi\nu\,(\text{mod}\;2\pi).
\end{equation}
Geometrically, $\nu$ counts how many times the unit Bloch vector $\hat{\mathbf n}_k$ winds around the origin in the chiral plane and remains invariant under continuous, gap-preserving deformations that maintain chiral symmetry. 
Longer-range intersublattice hopping allows multiple such windings as $k$ traverses the Brillouin zone.
Because the Zak--Berry phase is defined modulo $2\pi$, it retains only the parity of $\nu$, whereas the chiral winding number preserves its full integer value.

~\\

\noindent
\textbf{Uhlmann phase and topological kinks}

\noindent
For a system in thermal equilibrium with a heat bath at temperature $T$, the state at momentum $k$ is:
\begin{equation}\label{thermalstate}
\rho_k(T)=\frac{e^{-H_k/T}}{\operatorname{Tr}[e^{-H_k/T}]}=\frac{1}{2}\left[\mathbb I-r_k(T)\hat{\mathbf n}_k\cdot\boldsymbol{\sigma}\right].
\end{equation}
Here, $r_k(T)=\tanh(E_k/T)$, and the Boltzmann constant is set to unity.
Temperature therefore controls the statistical weights of the two bands without changing their eigenstates.

To characterize the geometry of these mixed states, we adopt the Uhlmann phase\,\cite{viyuela2014uhlmann,viyuela2014two,Huang2014}, which generalizes the Berry phase to density matrices. Each $\rho_k$ is represented by an amplitude $w_k=\sqrt{\rho_k}U_k$, where $U_k$ is unitary and $\rho_k=w_kw_k^\dagger$. The parallel-transport condition determines the evolution of $U_k$ along the momentum path, yielding a Uhlmann holonomy after one traversal of the Brillouin zone \,\cite{Uhlmann1986}. 
The full-loop Uhlmann phase is given by the argument of the overlap $\mathcal G_U(T)$ between the initial and final amplitudes:
\begin{equation}\label{eq.Uhlmann}
\Phi_U(T)=\arg\mathcal G_U(T)=\arg\operatorname{Tr} [w_{-\pi}^{\dagger}w_{\pi} ].
\end{equation}
For the chiral-symmetric two-band systems considered here, $\mathcal G_U(T)$ is real, and its sign determines whether $\Phi_U(T)=0$ or $\pi$. A temperature-driven transition occurs when the overlap changes sign across a critical temperature $T_c^{(j)}$ satisfying $\mathcal G_U(T_c^{(j)})=0$. The Uhlmann phase is undefined at the transition and switches between the two quantized values across it.

To resolve the transport along the momentum path rather than only at its endpoint, we introduce the partial overlap $\mathcal G_U(k,T)=\operatorname{Tr}[w_{-\pi}^{\dagger}(T)w_k(T)]$ and its phase $\Phi_U(k,T)=\arg \mathcal G_U(k,T)$. For the same planar Bloch trajectories, this overlap is also real. Its zeros define critical momenta $k_c^{(j)}(T)$, at which the initial and parallel-transported amplitudes are orthogonal in the Hilbert--Schmidt inner product and their relative phase is undefined. Crossing a simple zero reverses the sign of $\mathcal G_U(k,T)$, producing a $\pi$ phase jump---a topological kink in the Uhlmann holonomy\,\cite{viyuela2014uhlmann}.

Fig.~\ref{fig.system}(b) illustrates the thermal evolution of these kinks for representative systems with chiral winding numbers $\nu=1$, 2, and 3. 
At low temperature, the three systems exhibit one, two, and three topological kinks, respectively. 
Upon heating, the kinks move toward the Brillouin-zone boundary $k=\pi$ and disappear successively. A kink reaching this boundary makes the full-loop overlap vanish, marking a transition of $\Phi_U(T)$. Since each kink contributes a $\pi$ phase reversal along the momentum path, the endpoint phase retains only the parity of the number of kinks encountered. Consequently, thermal regimes with different numbers of remaining kinks can share the same $\Phi_U(T)$, despite having distinct momentum-resolved parallel-transport structures.

~\\
\noindent
\textbf{Uhlmann winding number}

\noindent
To convert the hierarchy revealed by the topological-kink structure into a global integer-valued characterization, we dimensionally extend the original 1D Hamiltonian by introducing an auxiliary polar coordinate $0\leq\vartheta\leq\pi$. 
With $\mathbf e^\perp=\mathbf e^i\times\mathbf e^j$ denoting the unit normal to the chiral plane, the extended Hamiltonian is written as\,\cite{Teo2010}:
\begin{equation}\label{DimensionalExtension}
H_k(\vartheta)
=
E_k\hat{\mathbf n}_k(\vartheta)
\cdot\boldsymbol{\sigma},
\quad
\hat{\mathbf n}_k(\vartheta)
=
\sin\vartheta\,\hat{\mathbf n}_k
+
\cos\vartheta\,\mathbf e^\perp.
\end{equation}
%
Because $\hat{\mathbf n}_k$ is orthogonal to $\mathbf e^\perp$, the extended Bloch vector remains normalized and the two bands retain the energies $\pm E_k$ for every $\vartheta$. The equator $\vartheta=\pi/2$ recovers the original 1D Hamiltonian, while the Bloch-vector directions at $\vartheta=0$ and $\pi$ are independent of momentum. 

At finite temperature, let $\widetilde{\mathcal G}_U(\vartheta,T)$ denote the complex Uhlmann overlap obtained by parallel transport around the momentum loop at fixed $\vartheta$. At the two poles, the density matrices along the momentum loop share a common eigenbasis, giving an identity Uhlmann holonomy and $\widetilde{\mathcal G}_U(0,T)=\widetilde{\mathcal G}_U(\pi,T)=1$.
Hence, varying $\vartheta$ from $0$ to $\pi$ produces a closed curve in the complex Uhlmann-overlap plane. 
Its equatorial point satisfies $\widetilde{\mathcal G}_U(\pi/2,T)=\mathcal G_U(T)$ and therefore has the phase $\Phi_U(T)$ whenever the overlap is nonzero. 
Because this equatorial overlap is real, it distinguishes only $\Phi_U=0$ and $\pi$. The full extended-overlap curve, however, can wind around the zero-overlap point multiple times. The number of windings defines the Uhlmann winding number:
\begin{equation}\label{UhlmannWinding}
\nu_U(T)
=
\frac{1}{2\pi i}
\int_0^\pi d\vartheta\,
\frac{
\partial_\vartheta
\widetilde{\mathcal G}_U(\vartheta,T)
}{
\widetilde{\mathcal G}_U(\vartheta,T)
}
\in\mathbb Z.
\end{equation}
This integer is well defined and quantized provided that $\widetilde{\mathcal G}_U(\vartheta,T)\neq0$ for every $\vartheta$, and can change only when the extended-overlap curve passes through the zero-overlap point.
In the pure-state limit, $\nu_U(T)\rightarrow\nu$ as $T\rightarrow0$, recovering the chiral winding number of the original 1D system (see Supplementary Note~1).

The phase-winding form of Eq.\,\eqref{UhlmannWinding} parallels that of the topological Uhlmann number defined for two-dimensional systems\,\cite{viyuela2014two,Viyuela2015}. 
In this construction, the full-loop Uhlmann phase is accumulated along one crystalline-momentum direction, and its winding is evaluated as the second momentum is varied. Here, $\vartheta$ is an auxiliary coordinate of the spectrum-preserving extension in Eq.\,\eqref{DimensionalExtension}. Equivalently, the overlap can be written as $\widetilde{\mathcal G}_U(\vartheta,T)=\operatorname{Tr}M(\vartheta,T)=\sum_jz_j(\vartheta,T)$, where $M$ denotes the product of Uhlmann amplitudes inside the trace in Eq.\,\eqref{eq.Uhlmann}, evaluated at fixed $\vartheta$, and $z_j$ are its complex eigenvalues. Thus, $\nu_U$ is the winding of the phase of their sum\,\cite{Huang2014} and can be evaluated from the reconstructed full-loop overlaps without tracking individual eigenvalue branches.

\begin{figure*}[t]
    \centering
    \includegraphics[width=0.98\linewidth]{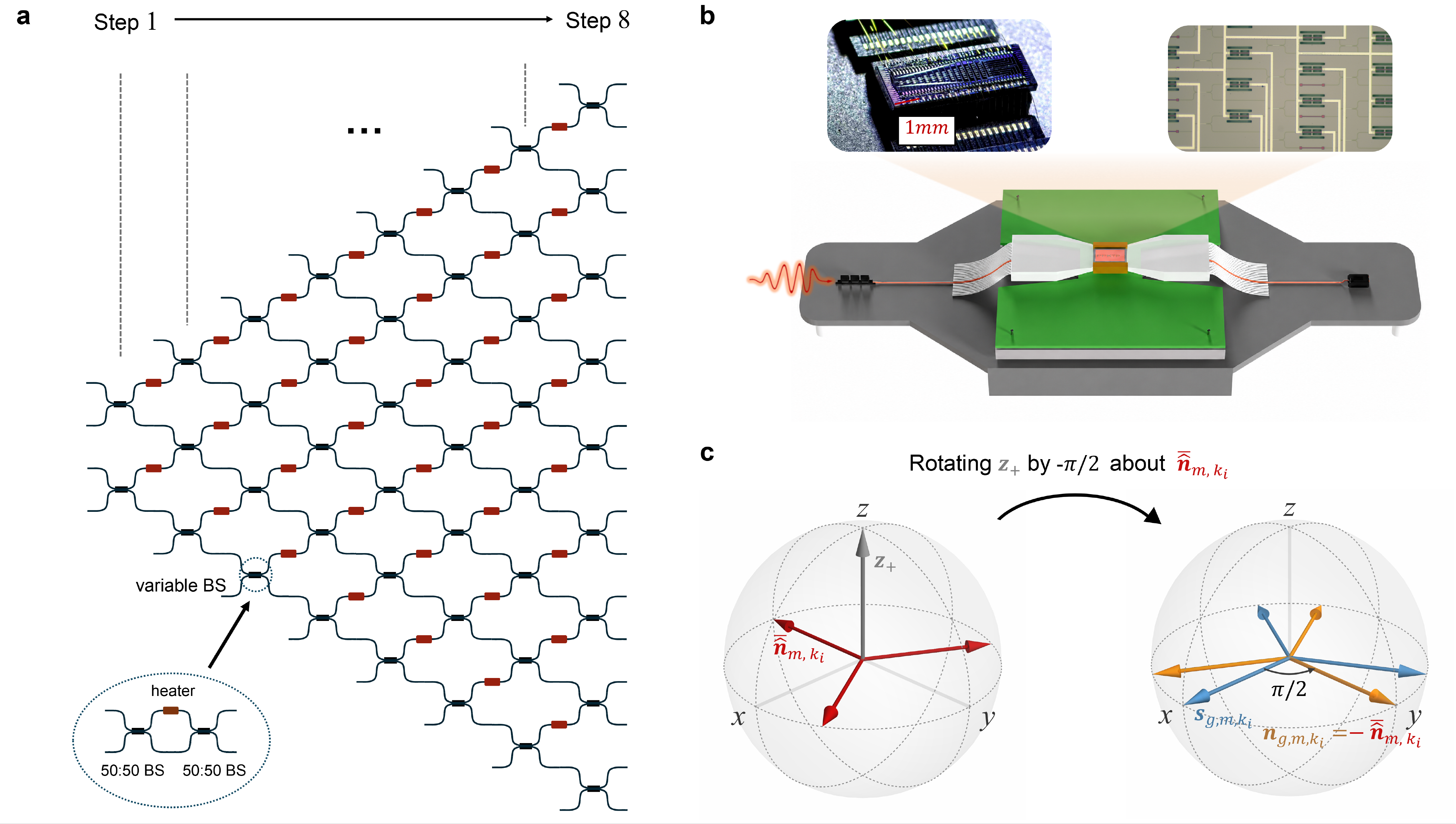}
    \caption{\textbf{Integrated photonic implementation of the quantum-walk protocol.} \textbf{a}, Schematic of the programmable waveguide circuit supporting up to eight single-site quantum-walk (QW) evolution layers, with selected layers indicated by dashed vertical lines. Programmable Mach–Zehnder interferometers and phase shifters implement the local coin operations, and fixed waveguide connections realize the conditional shifts. The inset shows a tunable beam splitter (BS) comprising two 50:50 BSs and a thermo-optic phase shifter.
    \textbf{b}, Schematic of the packaged photonic chip with input and output fibre arrays. The upper insets show a photograph of the fabricated chip (left; scale bar, 1 mm) and a magnified optical micrograph of the interferometric circuit (right). 
    \textbf{c}, Bloch-sphere illustration of band-state preparation using an auxiliary walk. The localized input \(|x=0,\uparrow\rangle\) has the same coin Bloch vector \(\mathbf{z}_{+}\) for all momentum components. The red vectors denote the momentum-dependent auxiliary rotation axes \(\overline{\hat{\mathbf{n}}}_{m,k_i}\) for representative momenta \(\{k_i\}\). The auxiliary operation \(\mathcal{U}_m(0,\pi/2)\) rotates \(\mathbf{z}_{+}\) by \(-\pi/2\) about each axis, producing the intermediate Bloch vectors \(\mathbf{s}_{g,m,k_i}\) (blue). A subsequent \(+\pi/2\) rotation about the \(z\) axis maps these vectors onto the ground-band Bloch vectors \(\mathbf{n}_{g,m,k_i}=-\overline{\hat{\mathbf{n}}}_{m,k_i}\) (orange). Applying the same sequence to the input \(|x=0,\downarrow\rangle\) prepares the corresponding excited-band states with Bloch vectors \(\mathbf{n}_{e,m,k_i}\).}
    
    \label{chip}
\end{figure*}

To determine how the Uhlmann winding number $\nu_U$ evolves with temperature, we consider the flat-band limit $E_k=E$ as an example. 
For this flat-band extension, all zeros of the extended Uhlmann overlap $\widetilde{\mathcal G}_U(\vartheta,T)$ are confined to the equator $\vartheta=\pi/2$. Hence, the critical temperatures at which $\nu_U$ can change are determined by:
\begin{equation}\label{UhlPhaseSimp}
\widetilde{\mathcal G}_U(\pi/2,T)
=
(-1)^\nu
\cos\!\left[
\pi\nu\operatorname{sech}(E/T)
\right].
\end{equation}
As the temperature increases, $\operatorname{sech}(E/T)$ increases monotonically from zero to one. 
For $\nu>0$, the equatorial overlap vanishes at $\nu$ critical temperatures:
\begin{equation}\label{CritT}
T_c^{(\ell)}
=
\frac{E}{
\operatorname{arcosh}[2\nu/(2\ell-1)]},
\quad
\ell=1,\ldots,\nu.
\end{equation}
Below the first critical temperature, the Uhlmann winding number remains equal to the chiral winding number, $\nu_U=\nu$. Upon crossing each critical temperature, $\nu_U$ decreases by one and eventually becomes zero above the highest critical temperature [Fig.~\ref{fig.system}(b)]. Since the equator recovers the original 1D Hamiltonian, $\widetilde{\mathcal G}_U(\pi/2,T)=\mathcal G_U(T)$. The Uhlmann phase and the Uhlmann winding number therefore change at the same critical temperatures in the flat-band limit (see Supplementary Notes~2 and 3). The corresponding results for dispersive SSH Hamiltonians are provided in Supplementary Note~4.

~\\
\noindent
\textbf{On-chip quantum-walk protocol}

\noindent
Here, we employ a discrete-time quantum walk (QW)\,\cite{Manouchehri2014,Su2019} to effectively realize the Bloch Hamiltonian $H_k$ in Eq.~\eqref{Hamiltonian}. The walk operator for one complete step is:
\begin{equation}\label{QWOperator}
\mathcal U_m(\theta_1,\theta_2)
=
S_-^{(m)}R(\theta_2)
S_+^{(m)}R(\theta_1),
\end{equation}
where the $m$-site conditional shifts are implemented by $m$ successive single-site shifts. The coin operator is $R(\theta)=i\sigma_x\exp(i\theta\sigma_y/2)$. 
The one-sided conditional shift $S_+^{(m)}$ moves the $\ket{\uparrow}$ component from $x$ to $x-m$ while leaving the $\ket{\downarrow}$ component unchanged. Conversely, $S_-^{(m)}$ moves the $\ket{\downarrow}$ component from $x$ to $x+m$ while leaving the $\ket{\uparrow}$ component unchanged. 
In momentum space, the walk operator defines the effective Bloch Hamiltonian through\,\cite{kitagawa2010exploring,Asboth2013}:
\begin{equation}\label{EffectiveQW}
\mathcal U_{m,k}
=
e^{-iH_k}
=
e^{-iE_k\hat{\mathbf n}_k\cdot\boldsymbol{\sigma}}.
\end{equation}
This relation determines the quasienergy $E_k$ and the unit Bloch vector $\hat{\mathbf n}_k$, whose explicit expressions are provided in Supplementary Notes~5. The $m$-site conditional shifts introduce the momentum dependence through $\sin(mk)$ and $\cos(mk)$. Consequently, in the topologically nontrivial regime, $\hat{\mathbf n}_k$ winds $m$ times as $k$ traverses the Brillouin zone, giving $|\nu|=m$. 
This construction synthesizes the effective long-range intersublattice coupling in the SSH model required for high winding numbers\,\cite{On2024Programmable}. 


\begin{figure*}[t]
    \centering
    \includegraphics[width=0.95\textwidth]{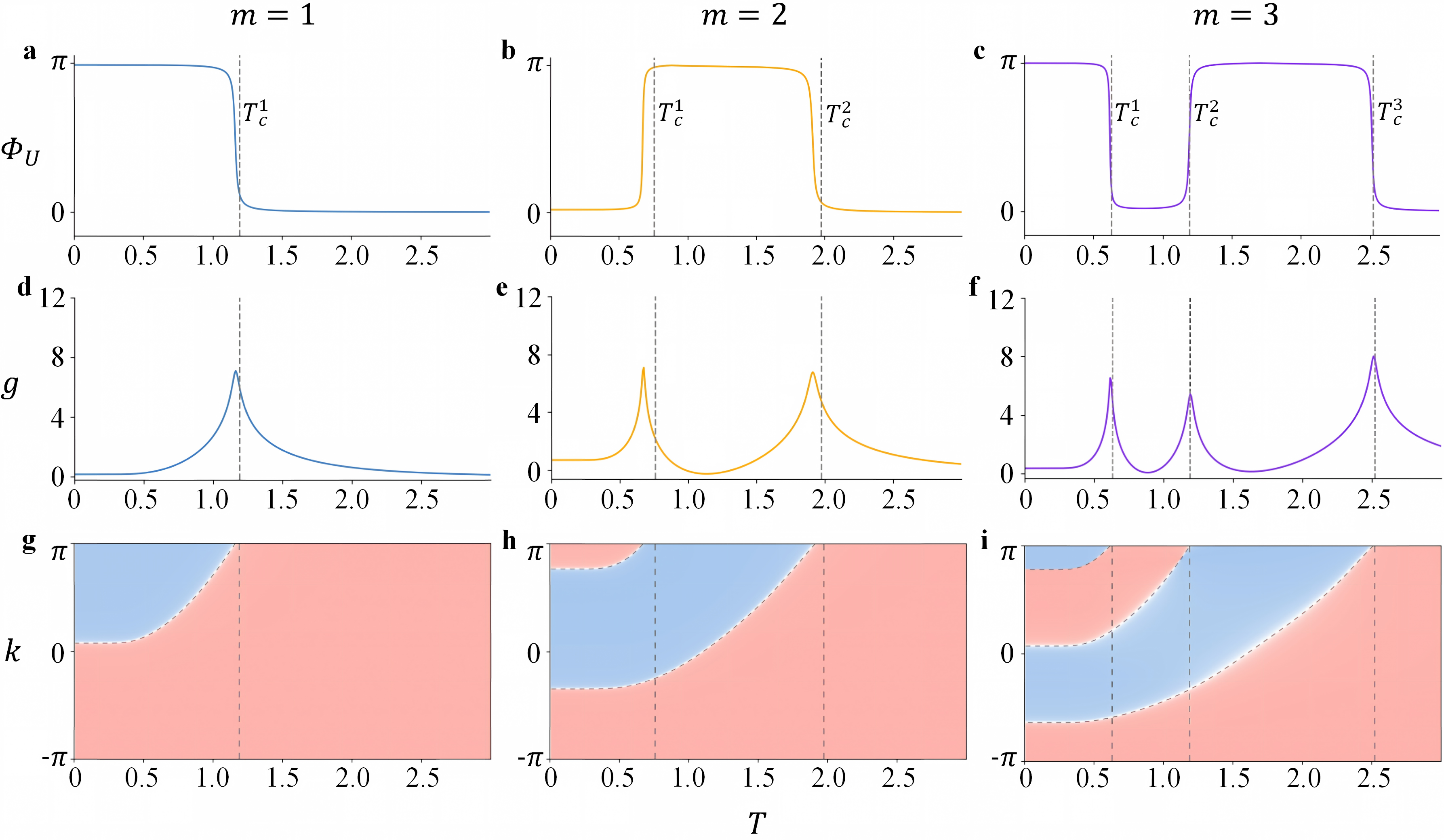}
    \caption{\textbf{Experimental observation of multiple temperature-driven Uhlmann transitions.} The three columns with $m=1$, $2$, $3$ correspond to systems with chiral winding numbers $\nu=1$, $2$, and $3$ respectively.  \textbf{a}--\textbf{c}, Reconstructed full-loop Uhlmann phase $\Phi_U(T)$, which switches between $0$ and $\pi$ once, twice, and three times, respectively.  \textbf{d}--\textbf{f}, Logarithmic Uhlmann overlap $g(T)$. The pronounced peaks indicate where the overlap approaches zero and identify the transition temperatures. 
    \textbf{g}--\textbf{i}, Reconstructed accumulated Uhlmann phase $\Phi_U(k,T)$ as a function of momentum and temperature. The red and blue regions correspond to phases $0$ and $\pi$, respectively. The gray dashed curves trace the theoretical phase-jump contours associated with the topological kinks of the Uhlmann holonomy. 
    %
    Each kink reaches the Brillouin-zone boundary at a critical temperature, accompanied by a switch of the full-loop Uhlmann phase. The vertical gray dashed lines indicate the theoretical critical temperatures. }
    \label{Transition}
\end{figure*}

The QW is implemented on a programmable silicon photonic chip\,\cite{Feng2022,Qiang2021,Zhu2024}. 
As shown in Fig.~\ref{chip}(a), the propagation direction through successive circuit layers encodes discrete time, while each position $x$ is represented by a pair of waveguides corresponding to the two coin states. At each evolution layer, the phase shifter inside  a Mach--Zehnder interferometer sets the effective splitting ratio of the variable beam splitter (BS), whereas the associated external phase shifter determines the relative phase between its two output modes. 
Together, the two phase shifters implement the local coin operation $R(\theta)$. 
The fixed waveguide connections between successive layers route the two coin components to neighboring position modes, thereby implementing the conditional shift. Independent control of the internal and external phase shifters allows the coin operation to be programmed separately at different positions and evolution layers.

%
The chip is interfaced with input and output fibre arrays [Fig.~\ref{chip}(b)]. It
integrates 44 interferometers constructed with multimode-interference couplers and thermo-optic phase shifters. 
Heralded single photons generated by spontaneous parametric down-conversion are injected through a selected channel of the input fibre array, and output photons are detected using a superconducting nanowire single-photon detector system. Calibrated thermo-optic phase shifters provide continuous phase tuning over $2\pi$, enabling the circuit to be programmed for state preparation and measurement.
Chip fabrication and calibration are described in Methods, with further calibration details and characterization results provided in Supplementary Notes~8.

The complete photonic circuit supports quantum walks of up to eight steps and is divided into two functional sections. The first section implements the quantum-walk sequences used for thermal state preparation. For the longest sequence with $m=3$, the output occupies an eight-dimensional coin--position space comprising four position sites $x\in\{0,1,2,3\}$ and two coin states $\{\ket{\uparrow},\ket{\downarrow}\}$ at each site.
The second section serves as a programmable high-dimensional measurement network. It applies unitary transformations that project the eight-mode output state onto different measurement bases before photon detection. Because quantum-walk dynamics are unitary, this network can also be programmed to implement inverse quantum-walk transformations that refocus selected states onto designated output modes\,\cite{karski2009quantum}. By varying the programmed transformation, the measurement network realizes the high-dimensional projections required for quantum-state tomography.

~\\
\noindent
\textbf{Preparation of thermal equilibrium states}

\noindent
To prepare the thermal equilibrium states, we choose
$(\theta_1,\theta_2)=(0,\pi)$ for the QWs with $m=1,2,$ and $3$.
At this working point, the corresponding effective Hamiltonians possess chiral symmetry with $\Gamma=\sigma_z$, have flat quasienergy bands with $E_k=E=\pi/2$, and realize the winding number $\nu=m$.
The flat spectrum makes the thermal occupation probabilities independent of momentum, allowing the same Boltzmann weights to be used for all momentum components and thereby facilitating the experimental preparation of the thermal equilibrium states.
The target thermal equilibrium state at each momentum is:
\begin{equation}\label{TargetThermalState}
\rho_k(T)
=
p_g(T)\ketbra{u_{g,k}}{u_{g,k}}
+
p_e(T)\ketbra{u_{e,k}}{u_{e,k}},
\end{equation}
where $\ket{u_{g,k}}$ and $\ket{u_{e,k}}$ are the eigenstates of $H_k$ with energies $-E$ and $+E$, respectively.
Their Boltzmann weights are
$p_{g/e}(T)=[1\pm\tanh(E/T)]/2$, where the plus and minus signs correspond to the ground and excited bands, respectively.

To prepare the ground- and excited-band eigenstates experimentally, we apply the auxiliary walk protocol\,\cite{xu2020}. We first initialize the walker in the localized position--coin states $\ket{x=0,\uparrow}$ and $\ket{x=0,\downarrow}$, respectively. In momentum space, each localized input corresponds to a uniform superposition over the Brillouin zone with a momentum-independent coin state.
We then apply the auxiliary walk operator $\mathcal U_m(0,\frac{\pi}{2})$ to the initial states. 
As shown in Fig.~\ref{chip}(c), the outputs are mapped to the target band eigenstates by a subsequent $+\pi/2$ rotation about the $z$ axis.
This rotation is implemented by the programmable phase shifters, yielding the ground- and excited-band eigenstates, respectively, at each momentum. 
Throughout our paper, the parameter $\theta_1$ of auxiliary walks is always fixed to zero. When the initial coin state is $\ket{\uparrow}$ or $\ket{\downarrow}$, the operation $S_{+}^{(m)}R(0)$ can be absorbed into the input state. We therefore implement only the $S_{-}^{(m)}R(\theta_2)$ part of $\mathcal U_m$ on the initial state. This reduces the number of walk steps and the possible position sites from seven, $\{-3,-2,-1,0,1,2,3\}$, to four, $\{0,1,2,3\}$. 

For each tomographic projection implemented by the programmable high-dimensional measurement network, the Boltzmann weights $p_{g/e}(T)$ are realized through the relative integration times of two successive measurement runs\,\cite{Wang2024Tomography}.
We first prepare and measure the ground state for an integration time proportional to $p_g(T)$ and then prepare and measure the excited state for an integration time proportional to $p_e(T)$.
The photon counts acquired for the same projection in the two runs are subsequently summed using the built-in processing function of the time-tagging software.
Because the ground and excited states are prepared in separate runs, the combined data represent an incoherent statistical mixture with the desired Boltzmann weights, corresponding to the thermal equilibrium state in Eq.~\eqref{TargetThermalState}.

We then apply threshold quantum-state tomography\,\cite{caruccio2025experimental}, which uses the measured diagonal populations to identify the off-diagonal density-matrix elements requiring characterization, thereby reducing the number of measurement settings.
The resulting photon counts are subsequently used in a maximum-likelihood reconstruction of the eight-dimensional position--coin density matrix, denoted by $\rho_{\mathrm{exp}}(T)$.
The momentum-resolved coin density matrix at each $k$ is obtained from the diagonal $k=k'$ block of its Fourier transform:
\begin{equation}\label{MomentumDensity}
\rho_k(T)
=
\mathcal N_k
\sum_{x,x'}
e^{+ik(x-x')}
\langle x|
\rho_{\mathrm{exp}}(T)
|x'\rangle,
\end{equation}
where $x,x'\in\{0,1,2,3\}$ and $\mathcal N_k$ is the normalization factor.
The resulting $\rho_k(T)$ is the experimental reconstruction of the thermal state defined in Eq.~\eqref{thermalstate}.

\begin{figure*}
    \centering
    \includegraphics[width=0.99\textwidth]{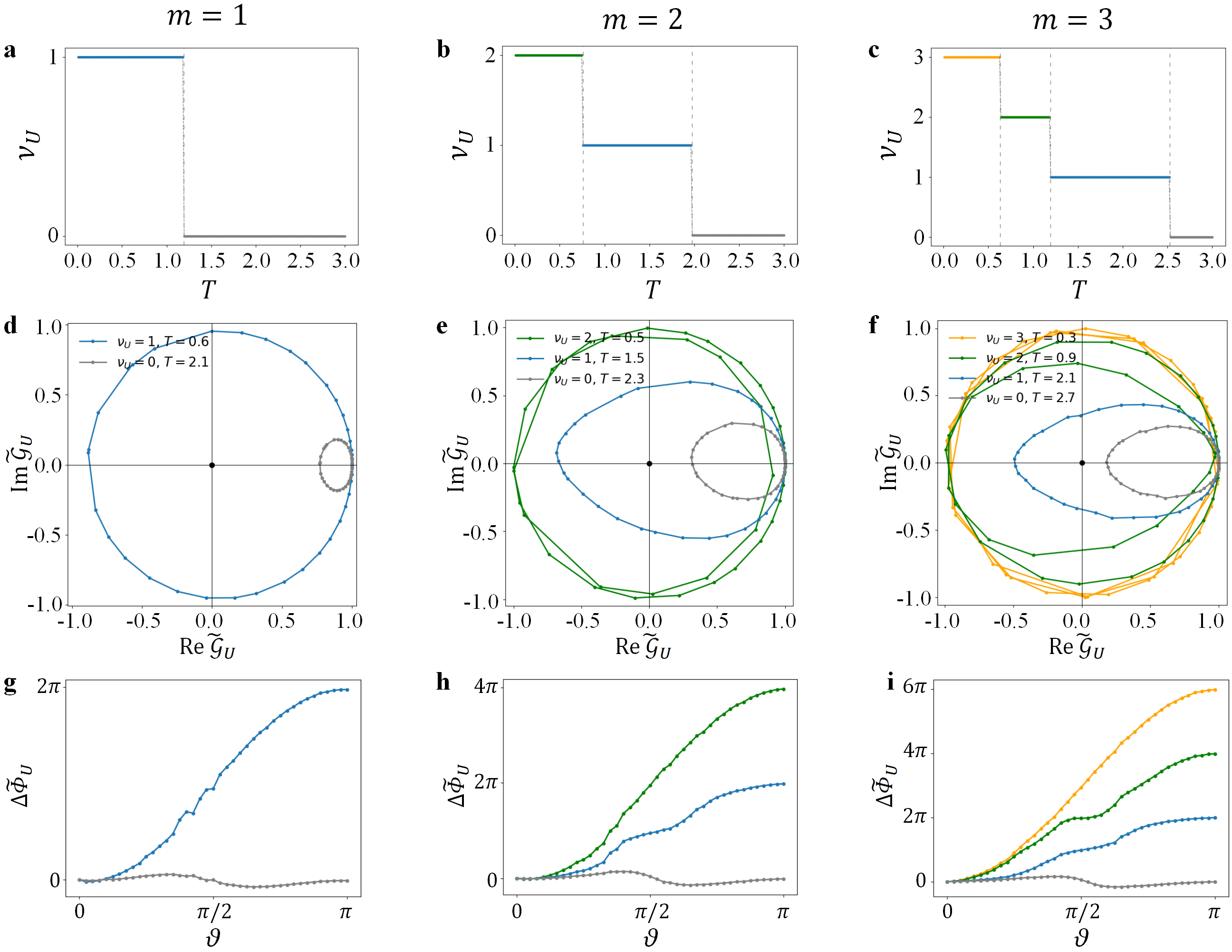}
    \caption{\textbf{Uhlmann winding numbers, extended Uhlmann-overlap curves, and continuously unwrapped phases.}
    \textbf{a}--\textbf{c}, Experimentally extracted Uhlmann winding number $\nu_U(T)$ as a function of temperature for $m=1$, $2$, and $3$, respectively. 
    The horizontal colored segments denote the sectors with $\nu_U=3$ (orange), $2$ (green), $1$ (blue), and $0$ (gray), while the vertical gray dashed lines mark the theoretical critical temperatures. 
    \textbf{d}--\textbf{f}, Closed trajectories of the extended Uhlmann overlap $\widetilde{\mathcal G}_U(\vartheta,T)$ in the complex plane at representative temperatures within the corresponding sectors. The trajectories are parametrized by the auxiliary coordinate from $\vartheta=0$ to $\pi$, and the temperatures and winding numbers are specified in the legends. The larger black dot marks the zero-overlap point at the origin, whereas the smaller black dot at $(1,0)$ denotes the common starting and ending point of the trajectories.
    \textbf{g}--\textbf{i}, Continuously unwrapped phase change $\Delta\widetilde{\Phi}_U(\vartheta,T)$ associated with the trajectories in (d)--(f). The phase is referenced to zero at $\vartheta=0$, and the same representative temperatures and color convention are used as in (d)--(f).}
    \label{UWN}
\end{figure*}

~\\
\noindent
\textbf{Observation of multiple thermal transitions}

\noindent
Using the reconstructed momentum-resolved density matrices $\rho_k(T)$, we construct the Uhlmann holonomy around the Brillouin zone from the parallel-transport unitaries connecting adjacent density matrices.
This approach extends the use of tomography for topological characterization in photonic systems\,\cite{Xu2018,Barkhofen2018,Asapanna2025,Lin2026} from pure states to thermal mixed states.
For each temperature, the momentum loop is discretized into $N=100$ equally spaced points. The parallel-transport unitary $V_i(T)$ between each pair of adjacent density matrices is obtained from a polar decomposition, including the link that closes the loop. Their ordered product gives the Uhlmann holonomy $U_{\mathrm{Uhl}}(T)=V_N(T)\cdots V_1(T)$. The corresponding Uhlmann overlap is $\mathcal G_U(T)=\operatorname{Tr}[\rho_{-\pi}(T)U_{\mathrm{Uhl}}(T)]$, and the Uhlmann phase is $\Phi_U(T)=\arg\mathcal G_U(T)$ [see Supplementary Notes~7.2 for details]. 

The resulting Uhlmann phases are shown in Figs.~\ref{Transition}(a)--(c). The Uhlmann phase switches between $0$ and $\pi$ whenever the real overlap $\mathcal G_U(T)$ crosses zero and changes sign. This produces the sequences $\pi\rightarrow0$, $0\rightarrow\pi\rightarrow0$, and $\pi\rightarrow0\rightarrow\pi\rightarrow0$ for $\nu=1$, $2$, and $3$, respectively. 
Thus, for the flat-band Hamiltonians realized here, the number of successive Uhlmann-phase transitions equals the zero-temperature winding number $\nu$.
The observed phase switches agree with the theoretical critical temperatures indicated by the gray dashed lines.
To provide a complementary experimental signature of the thermal transitions, we examine the logarithmic Uhlmann overlap, $g(T)=-\ln|\mathcal G_U(T)|^2$, in analogy with the rate function for Uhlmann quenches\,\cite{Tang2024Uhlmann}.
In the ideal model, $\mathcal G_U(T)$ becomes zero at each critical temperature, at which the Uhlmann phase is ill-defined and $g(T)$ diverges. 
Experimentally, the reduced overlap magnitude is reflected in the pronounced peaks shown in Figs.~\ref{Transition}(d)--(f).
The peak positions locate local minima of the overlap magnitude and are consistent with both the observed Uhlmann-phase switches and the theoretical critical temperatures.

Although the full-loop Uhlmann phase and the logarithmic overlap identify the transition temperatures, they do not reveal how each transition develops along the Brillouin zone.
Taking $k=-\pi$ as the starting point of the Uhlmann parallel transport, we evaluate the phase accumulated along the momentum path from $-\pi$ to each intermediate momentum $k$.
For this open momentum segment, we define the partial Uhlmann overlap $\mathcal G_U(k,T)=\operatorname{Tr}[w_{-\pi}^{\dagger}(T)w_k(T)]$ and its accumulated phase $\Phi_U(k,T)=\arg\mathcal G_U(k,T)$, where the amplitudes $w_k(T)$ satisfy the same Uhlmann parallel-transport condition.
Upon reaching $k=\pi$, the momentum path closes and the accumulated phase gives the full-loop Uhlmann phase $\Phi_U(T)$.
Figs.~\ref{Transition}(g)--(i) show the reconstructed accumulated phases $\Phi_U(k,T)$ for systems with $\nu=1$, $2$, and $3$, respectively. The red and blue regions correspond to $\Phi_U(k,T)=0$ and $\pi$, while the curved dashed contours mark the discontinuous phase changes. At low temperature, the three systems exhibit one, two, and three such contours, respectively. As the temperature increases, these contours reach $k=\pi$ successively at the critical temperatures indicated by the vertical gray dashed lines and subsequently disappear from the Brillouin-zone path.

Each curved contour has a direct geometric meaning. At a fixed temperature, each contour identifies a critical momentum $k_c^{(j)}(T)$ at which $\mathcal G_U(k,T)=0$. The initial amplitude and the amplitude parallel transported to this momentum are therefore orthogonal, making their relative phase ill-defined. Crossing the critical momentum reverses the sign of the overlap and changes $\Phi_U(k,T)$ by $\pi$. Such a singular phase reversal constitutes a topological kink in the Uhlmann holonomy\,\cite{viyuela2014uhlmann}. 
When a kink reaches $k=\pi$, the full-loop overlap becomes zero. 
Crossing the corresponding critical temperature reverses its sign and switches the full-loop Uhlmann phase.
This momentum-resolved picture shows that identical full-loop Uhlmann phases can arise from distinct Uhlmann parallel-transport processes.
Because $\Phi_U(T)$ retains only the phase at the endpoint of the momentum loop, it distinguishes only whether the number of kinks encountered along the path is even or odd.
The kink structure therefore resolves these successive changes locally along momentum, while the Uhlmann winding number observed below provides a global integer-valued characterization of the same parallel-transport process.

~\\
\noindent
\textbf{Measurement of integer Uhlmann winding numbers}

\noindent
To obtain a global integer-valued characterization, we extend the auxiliary-walk preparation to the ground- and excited-band eigenstate textures of the dimensionally extended Hamiltonian $H_k(\vartheta)$ and use them to construct the corresponding thermal mixed states.
Because the dimensional extension preserves the flat quasienergy $E=\pi/2$, the Boltzmann weights remain independent of both $k$ and $\vartheta$.
Varying $\vartheta$ therefore changes the prepared band eigenstates, while their statistical weights remain fixed at a given temperature.

For each of the three QW configurations with $m=1$, $2$, and $3$, we divide the interval $0\leq\vartheta\leq\pi$ into 20 equal segments, giving 21 sampled values that include the two poles and the equator.
At each $\vartheta$, the walker is initialized in $\ket{x=0,\uparrow}$ and $\ket{x=0,\downarrow}$.
Applying the auxiliary walk operator $\mathcal U_m(0,\pi-\vartheta)$ followed by a $\pi/2$ rotation about the $z$ axis maps the two inputs onto the ground- and excited-band eigenstate textures of $H_k(\vartheta)$, respectively.
The successive one-sided conditional shifts generate the required $mk$ dependence, while the second coin angle controls the polar angle of the extended Bloch-vector texture.
For each $\vartheta$ and temperature $T$, the two band outputs are combined incoherently by setting their relative integration times according to $p_g(T)$ and $p_e(T)$, yielding the extended thermal states $\rho_k(\vartheta,T)$.

Quantum-state tomography is performed for each sampled $\vartheta$ and temperature $T$.
%
We then Fourier transform the position-space blocks of the reconstructed position--coin density matrix to obtain the momentum-resolved density matrices $\rho_k(\vartheta,T)$ along the Brillouin zone.
For each fixed $\vartheta$, the Uhlmann holonomy is then constructed from adjacent density matrices along the discretized momentum loop, yielding the complex full-loop overlap $\widetilde{\mathcal G}_U(\vartheta,T)$.
%
The Uhlmann winding number is then evaluated from the ordered set of sampled overlaps using the discrete form of Eq.~\eqref{UhlmannWinding} (see Supplementary Note~7.3).
%
Fig.~\ref{UWN}(a)--(c) show the resulting temperature dependence of $\nu_U$ for $m=1$, $2$, and $3$. At low temperature, the winding number takes its pure-state value $\nu_U=\nu$. As the temperature increases, it passes successively through the lower integer sectors and reaches $\nu_U=0$ above the highest critical temperature. The transitions occur at the same critical temperatures as the switches of the full-loop Uhlmann phase in Figs.~\ref{Transition}(a)--(c).

Geometrically, the Uhlmann winding number counts how many times this closed trajectory encircles the zero-overlap point. At a critical temperature, the curve passes through this point, making both its phase and winding number ill-defined.
The corresponding complex overlap curves at representative temperatures are shown in Fig.~\ref{UWN}(d)--(f). The curves in the $\nu_U=3$, $2$, and $1$ sectors encircle the zero-overlap point three, two, and one times, respectively, whereas those in the $\nu_U=0$ sector do not encircle it.
Fig.~\ref{UWN}(g)--(i) show the accumulated, continuously unwrapped phases of the same overlap trajectories.
We denote this accumulated phase change by $\Delta\widetilde{\Phi}_U(\vartheta,T)$ and set $\Delta\widetilde{\Phi}_U(0,T)=0$.
Its endpoint value directly gives the Uhlmann winding number through $\Delta\widetilde{\Phi}_U(\pi,T)=2\pi\nu_U(T)$.
Accordingly, a trajectory at temperature $T$ accumulates a total phase of $2\pi\nu_U(T)$, whereas in the sector with zero Uhlmann winding, the accumulated phase can vary at intermediate values of $\vartheta$ but returns to zero at the end of the scan.
Taken together, these two representations encode the same integer invariant geometrically through the encirclement of the zero-overlap point and through the total phase accumulated along $\vartheta$.
Whereas the full-loop Uhlmann phase retains only the parity of this integer, this integer invariant resolves the distinct intermediate thermal sectors that can share the same value of $\Phi_U$.


~\\
\noindent
{\bf\large Discussion}

\noindent
In conclusion, we have used an integrated photonic chip to experimentally extract integer-valued Uhlmann winding numbers for thermal states of chiral quantum-walk Hamiltonians and track their evolution with temperature.
The geometric information not fully captured by the two-valued topological Uhlmann phase appears locally as singular phase reversals of the partial overlap along the physical momentum path, known as topological kinks, and globally as the integer winding of the closed extended-overlap trajectory.
Within the spectrum-preserving extension of the flat-band Hamiltonians realized here, the global winding numbers distinguish thermal regimes that share the same value of $\Phi_U$, whereas the Uhlmann phase retains only the parity of the integer winding.

The spectrum-preserving extension uses the normal to the chiral plane to tilt the Bloch vectors without introducing additional azimuthal winding, while retaining the original thermal populations at each momentum. The original 1D loop is recovered at the equator, and the two poles provide reference loops with trivial Uhlmann transport, closing the overlap trajectory. This auxiliary family is constructed from the original Hamiltonian and its chiral operator, without requiring an independently specified higher-dimensional model. Nevertheless, the resulting phase-winding expression has the same mathematical structure as the Uhlmann number defined for two-dimensional systems\,\cite{viyuela2014two,Viyuela2015}. 
Changes in this winding occur at zeros of the original full-loop overlap, with no additional transitions away from the equator. 
The extension thus provides an integer characterization of thermal sectors that share the same full-loop Uhlmann phase.
%
We anticipate that extending this construction to nonthermal mixed states\,\cite{Hu2016} and multiband Uhlmann processes will clarify how non-Gibbs populations and coherences, as well as multiband holonomy, reshape the integer structure of Uhlmann geometry.

~\\
\noindent
{\bf\large Methods}

\noindent
\textbf{Chip fabrication and calibration}

\noindent
The photonic chip was fabricated on a standard 220-nm silicon-on-insulator platform using a CMOS-compatible process. The circuit contains 88 multimode-interference couplers, together with 79 thermo-optic phase shifters, 80 electrical contact pads, and 44 input and output edge couplers. The phase shifters were driven by voltages ranging from 0 to 1 V, corresponding to heater currents of up to approximately 2.5 mA. This range provides continuous phase tuning over $2\pi$. 

The thermo-optic phase shifters are divided into two types. Internal phase shifters controlled the effective splitting ratios of the MZIs, allowing them to function as variable beam splitters. External phase shifters, located between successive MZIs, controlled the relative phases between the optical modes. All phase shifters were calibrated before programming the quantum-walk circuit.

The internal phase shifters were calibrated sequentially along the diagonal paths of the MZI mesh in Fig.~\ref{chip}(a), proceeding from right to left along each path before moving to the next diagonal. To calibrate the external phase shifters, neighboring MZIs were programmed to form a larger effective MZI with the target phase shifter in one arm, converting phase shifts into measurable intensity variations. The two outer MZIs acted as balanced couplers to split and recombine the light, while the two intermediate MZIs routed light through the interferometer arms.

In both cases, sweeping the target heater current produced an output interference fringe, from which the phase--current response of the target phase shifter and the extinction ratio of the corresponding interferometer were extracted. The 36 characterized MZIs exhibited a mean extinction ratio of $38.17$~dB with a standard deviation of $3.21$~dB. The 21 effective interferometers used to characterize the external phase shifters exhibited a mean extinction ratio of $30.73$~dB with a standard deviation of $3.88$~dB. These results indicate high interferometric contrast in the characterized circuits. 

~\\
\noindent
{\bf\large Acknowledgements}

\noindent
We thank Bing Chen for helpful discussions. This work was supported by National Natural Science Foundation of China (Nos. 12574401, 12474489, 12474494, 62675079); the Guangdong Project (Grant No. 2024TQ08A680); the Guangdong
Provincial Quantum Science Strategic Initiative (Grant
Nos. GDZX2603007, GDZX2403005, GDZX2403001, GDZX2203001, 
GDZX2403002,  GDZX2506001, GDZX2503001, and GDZX2503002);
Guangdong Basic and Applied Basic Research Foundation (Grant No. 2024A1515012405);
Shenzhen Fundamental Research Program (Grant No. JCYJ20240813153139050);
the CAS Project for Young Scientists in Basic Research (YSBR-131).

~\\
\noindent
{\bf\large Author contributions}

\noindent
B.Y., Q.-Q.W., and H.H. contributed equally to this work.
Q.-Q.W. and Y.-Y.Z. conceived the experimental protocol.
B.Y. and H.H. performed the experiment with help from Q.-Q.W., B.-T.Y., X.-Y.X. and Y.-Y.Z.  
F.G. designed the quantum walk chip. 
B.Y. and Q.-Q.W. analyzed the experimental data. 
B.Y., Q.-Q.W., and Y.-Y.Z. wrote the manuscript with input from all authors.
All authors discussed the results and commented on the manuscript.

~\\
\noindent
{\bf\large Competing interests}

\noindent
The authors declare no competing financial or non-financial interests.

~\\
\noindent
{\bf\large Data availability}

\noindent
Source data are provided with this paper.



\bibliographystyle{apsrev4-1PRX.bst}
\bibliography{ref}


\section*{Supplemental Materials}

\renewcommand{\theequation}{S\arabic{equation}}
\setcounter{equation}{0}

\suppnote{Quantization and pure-state limit of the Uhlmann winding number}
\label{sec:uw}

In this section, we establish the integer quantization of $\nu_U$ and show that it reduces to the chiral winding number in the pure-state limit.
The extended Hamiltonian $H_k(\vartheta)$ reduces to the original Hamiltonian $H_k$ at the equator. 
At $\vartheta=0$ and $\pi$, the unit Bloch vector points along $\mathbf e^\perp$ and $-\mathbf e^\perp$, respectively, for every momentum $k$.
The density matrices along each polar momentum loop therefore share a common eigenbasis. For a dispersive spectrum, their eigenvalues may still depend on momentum through $E_k$. However, $\sqrt{\rho_k}$ and its momentum derivative remain diagonal in the same basis, so their commutator is zero. The Uhlmann connections at the two poles are consequently zero, and the corresponding holonomies are the identity. Together with momentum periodicity and normalization, this gives:
\begin{equation}\label{eq:uw_poles}
\widetilde{\mathcal G}_U(0,T)
=
\widetilde{\mathcal G}_U(\pi,T)
=
1.
\end{equation}
Therefore, varying $\vartheta$ from $0$ to $\pi$ generates a closed curve in the complex Uhlmann-overlap plane.

To demonstrate the integer quantization, we consider a temperature away from an overlap singularity, so that $\widetilde{\mathcal G}_U(\vartheta,T)\neq0$ for every $\vartheta$. We write $\widetilde{\mathcal G}_U(\vartheta,T)=R(\vartheta,T)e^{i\phi(\vartheta,T)}$, where $R>0$ and $\phi$ is continuously unwrapped along the curve. 
Substitution into the definition of the Uhlmann winding number in Eq.\,(7) of the main text gives a modulus contribution proportional to $\ln[R(\pi,T)/R(0,T)]$ and a phase contribution. 
Eq.\,\eqref{eq:uw_poles} gives $R(0,T)=R(\pi,T)=1$, so the modulus contribution is zero. 
Since the initial and final overlaps are both equal to unity, their continuously unwrapped phases can differ only by an integer multiple of $2\pi$. Therefore:
\begin{equation}\label{eq:uw_quant}
\nu_U(T)
=
\frac{\phi(\pi,T)-\phi(0,T)}{2\pi}
\in\mathbb Z.
\end{equation}
The integer $\nu_U$ remains unchanged under any continuous deformation for which the overlap remains nonzero. It can change only when $\widetilde{\mathcal G}_U(\vartheta,T)=0$, at which point both the overlap phase and $\nu_U$ become undefined.

For a gapped system, the density matrices approach the occupied-band projectors in the zero-temperature limit. Uhlmann parallel transport then reduces to Berry parallel transport, giving $\lim_{T\rightarrow0}\widetilde{\mathcal G}_U(\vartheta,T)=e^{i\Phi_B(\vartheta)}$. Choosing the continuously unwrapped Berry phase with $\Phi_B(0)=0$, the Berry phase along the momentum loop at fixed $\vartheta$ is:
\begin{equation}\label{eq:uw_berry}
\Phi_B(\vartheta)
=
\frac{1-\cos\vartheta}{2}
\oint_{\mathrm{BZ}}d\alpha_k
=
\pi\nu(1-\cos\vartheta).
\end{equation}
The Berry phase therefore changes by $2\pi\nu$ between the two poles. Eq.\,\eqref{eq:uw_quant} then gives:
\begin{equation}\label{eq:uw_zeroT}
\lim_{T\rightarrow0}\nu_U(T)
=
\frac{\Phi_B(\pi)-\Phi_B(0)}{2\pi}
=
\nu.
\end{equation}
Thus, the Uhlmann winding number recovers the chiral winding number of the original 1D Hamiltonian in the pure-state limit.
In the infinite-temperature limit, $\rho_k(\vartheta,T)$ approaches $\mathbb I/2$ for every $k$ and $\vartheta$. The Uhlmann connection is therefore zero, and the corresponding holonomy is the identity. The overlap curve reduces to the point $\widetilde{\mathcal G}_U=1$, giving $\lim_{T\rightarrow\infty}\nu_U(T)=0$.

\suppnote{Analytical solution for the flat-band extension}
\label{sec:flatU}

We derive the Uhlmann overlap for the flat-band extension with $E_k=E$ and $\nu>0$. Writing $s=\sin\vartheta$ and $c=\cos\vartheta$, the extended unit Bloch vector is:
\begin{equation}\label{eq:flat_n}
\hat{\mathbf n}_k(\vartheta)
=
s\left(
\cos\alpha_k\,\mathbf e^i
+
\sin\alpha_k\,\mathbf e^j
\right)
+
c\,\mathbf e^\perp,
\end{equation}
where $\oint_{\mathrm{BZ}}d\alpha_k=2\pi\nu$. The thermal-state Bloch radius $r(T)=\tanh(E/T)$ is independent of momentum and $\vartheta$.

For a two-level thermal state, the Uhlmann connection takes the form\,\cite{viyuela2014uhlmann,morachis2021topological}:
\begin{equation}\label{eq:flat_A}
A_U
=
-\frac{i}{2}f(T)
\left(
\hat{\mathbf n}\times d\hat{\mathbf n}
\right)\cdot\boldsymbol{\sigma},
\qquad
f(T)
=
1-\operatorname{sech}(E/T).
\end{equation}
At fixed $\vartheta$, the generator in Eq.\,\eqref{eq:flat_A} rotates with $\alpha_k$. 
We choose the angular origin such that $\alpha_{-\pi}=0$ and write $\sigma_\perp=\mathbf e^\perp\cdot\boldsymbol{\sigma}$. The co-rotating transformation $R(\alpha_k)=\exp[-i\alpha_k\sigma_\perp/2]$ then removes this dependence.
After one traversal of the Brillouin zone, the resulting Uhlmann holonomy is:
\begin{align}\label{eq:flat_hol}
U_{\mathrm{Uhl}}(\vartheta,T)
&=
(-1)^\nu
\exp\!\left[
-i\pi\nu\,
\mathbf h\cdot\boldsymbol{\sigma}
\right],
\nonumber\\
\mathbf h
&=
-fsc\,\mathbf e^i
+
\left(fs^2-1\right)\mathbf e^\perp.
\end{align}
The norm of this effective rotation axis is:
\begin{equation}\label{eq:flat_C}
C(\vartheta,T)
=
|\mathbf h|
=
\sqrt{
1-r^2(T)\sin^2\vartheta
}.
\end{equation}

Evaluating $\widetilde{\mathcal G}_U=\operatorname{Tr}[\rho_{-\pi}U_{\mathrm{Uhl}}]$ gives the complete Uhlmann overlap:
\begin{align}\label{eq:flat_G}
\widetilde{\mathcal G}_U(\vartheta,T)
=
(-1)^\nu
\Bigg[
&\cos\!\left(\pi\nu C\right)
\nonumber\\
&-
i\frac{r(T)\cos\vartheta}{C}
\sin\!\left(\pi\nu C\right)
\Bigg].
\end{align}
This expression gives $\widetilde{\mathcal G}_U=1$ at both poles and reproduces the pure-state and infinite-temperature limits discussed in Supplementary Note\,\ref{sec:uw}.

The possible zeros of the overlap follow directly from Eq.\,\eqref{eq:flat_G}. At finite temperature and $\vartheta\neq\pi/2$, vanishing of the imaginary part requires $\sin(\pi\nu C)=0$. Under this condition, $\cos(\pi\nu C)=\pm1$, and the real part remains nonzero. The zero-overlap condition is therefore restricted to the equator. Setting $\vartheta=\pi/2$ gives:
\begin{equation}\label{eq:flat_equator}
\widetilde{\mathcal G}_U(\pi/2,T)
=
(-1)^\nu
\cos\!\left[
\pi\nu\operatorname{sech}(E/T)
\right].
\end{equation}
The corresponding critical temperatures are therefore:
\begin{equation}\label{eq:flat_Tc}
T_c^{(\ell)}
=
\frac{E}{
\operatorname{arcosh}
\!\left[
2\nu/(2\ell-1)
\right]},
\qquad
\ell=1,\ldots,\nu.
\end{equation}

Each zero in Eq.~\eqref{eq:flat_equator} is simple and changes the winding of the complex overlap curve by one. Since $\nu_U$ approaches $\nu$ at zero temperature and $0$ at infinite temperature, it decreases successively across the critical temperatures. Away from the transition points, $\nu_U=\nu-\ell$ for $T_c^{(\ell)}<T<T_c^{(\ell+1)}$, where $T_c^{(0)}=0$ and $T_c^{(\nu+1)}=\infty$. 

\suppnote{Critical temperatures of the Uhlmann phase}
\label{sec:TcU}

For the chiral two-band systems considered here, the Uhlmann connection at each momentum is proportional to the same generator $\mathbf e^\perp\cdot\boldsymbol{\sigma}$, where $\mathbf e^\perp$ is normal to the chiral plane. Consequently, the connections at different momenta commute, and the Uhlmann holonomy can be evaluated without path ordering, yielding\,\cite{viyuela2014uhlmann}:
\begin{equation}\label{eq:GUd}
\mathcal G_U(T)
=
(-1)^\nu
\cos\!\left[
\frac{1}{2}
\oint_{\mathrm{BZ}}
\operatorname{sech}\!\left(\frac{E_k}{T}\right)
d\alpha_k
\right].
\end{equation}
The Uhlmann overlap $\mathcal G_U(T)$ is real, so its sign determines the Uhlmann phase. At a critical temperature, $\mathcal G_U(T)=0$ and the Uhlmann phase becomes ill-defined. Crossing this temperature changes the sign of $\mathcal G_U(T)$, thereby switching the Uhlmann phase between $0$ and $\pi$.

For a dispersive system with a positive winding number and a monotonically increasing unwrapped angle $\alpha_k$, the critical temperatures are determined by:
\begin{equation}\label{eq:Tcd}
\oint_{\mathrm{BZ}}
\operatorname{sech}\!\left(
\frac{E_k}{T_c^{(\ell)}}
\right)
d\alpha_k
=
(2\ell-1)\pi,
\end{equation}
where $\ell=1,\ldots,\nu$ for $\nu>0$. 
%
Consequently, the accumulated angle on the left-hand side of Eq.\,\eqref{eq:Tcd} increases continuously from zero to $2\pi\nu$ with temperature, giving $\nu$ distinct critical temperatures.
Let $E_{\min}=\min_{k\in\mathrm{BZ}}E_k$ and $E_{\max}=\max_{k\in\mathrm{BZ}}E_k$. For a monotonic winding, the critical temperatures satisfy:
\begin{equation}\label{eq:Tcb}
\begin{aligned}
T_c^{(\ell)}
&\geq
\frac{E_{\min}}{
\operatorname{arcosh}
\!\left[2\nu/(2\ell-1)\right]},
\\
T_c^{(\ell)}
&\leq
\frac{E_{\max}}{
\operatorname{arcosh}
\!\left[2\nu/(2\ell-1)\right]}.
\end{aligned}
\end{equation}
In the flat-band limit $E_k=E$, the momentum dependence drops out and the critical temperatures reduce to:
\begin{equation}\label{eq:Tcf}
T_c^{(\ell)}
=
\frac{E}{
\operatorname{arcosh}
\!\left[2\nu/(2\ell-1)\right]}.
\end{equation}
Thus, in the flat-band limit, the Uhlmann phase and the Uhlmann winding number undergo transitions at the same critical temperatures.

\suppnote{Critical-temperature comparison for dispersive SSH models}
\label{sec:dispTc}

The flat-band results in Supplementary Notes\,\ref{sec:flatU} and \ref{sec:TcU} establish analytically that the Uhlmann phase and the Uhlmann winding number change at the same critical temperatures. 
To examine this correspondence for dispersive bands, we consider a generalized SSH model as a concrete example\,\cite{chen2020elementary}:
\begin{equation}\label{eq:dispSSH}
H_{\nu,k}
=
[t_0+t_\nu\cos(\nu k)]\sigma_x
+
t_\nu\sin(\nu k)\sigma_y ,
\end{equation}
where $t_0$ is the intracell hopping and $t_\nu$ denotes intersublattice hopping across $\nu$ unit cells. For $|t_0|<|t_\nu|$, the chiral winding number is $\nu$, while the positive-band energy is $E_{\nu,k}=[t_0^2+t_\nu^2+2t_0t_\nu\cos(\nu k)]^{1/2}$. The spectrum is dispersive except at $t_0=0$.

We set $t_\nu=1$ and sample 21 values of $\lambda=t_0/t_\nu$ uniformly over $-0.85\leq\lambda\leq0.85$ for $\nu=1,2,3$. At each parameter point, the critical temperatures of the Uhlmann phase are obtained from the zeros of the physical overlap $\mathcal G_U(T)$. 
Independently, the Uhlmann winding number is calculated from the complete extended overlap curve $\widetilde{\mathcal G}_U(\vartheta,T)$. 
The momentum loop is discretized into 101 points. The critical temperatures of the Uhlmann phase are obtained by bracketing sign changes of $\mathcal G_U(T)$ and solving $\mathcal G_U(T)=0$ by bisection. Independently, the auxiliary interval $0\leq\vartheta\leq\pi$ is sampled at 51 points, and each temperature interval in which the winding of the complete overlap curve changes is refined by bisection.

As shown in Supplementary Fig.~\ref{Compare}, the critical temperatures obtained independently from the two methods are in good agreement.
%
At every winding transition identified in the scan, the minimum of $|\widetilde{\mathcal G}_U(\vartheta,T)|$ is attained at the equator and approaches zero within numerical accuracy. Since $\widetilde{\mathcal G}_U(\pi/2,T)=\mathcal G_U(T)$, each change in $\nu_U$ coincides with a zero of the Uhlmann overlap of the original SSH Hamiltonian. 
This numerical result supports the correspondence for the sampled dispersive SSH models.
The small residual offsets from the diagonal arise from finite momentum discretization and the numerical tolerance used to locate the winding transitions.
We note that increasing the number of sampled values of $\lambda$ leaves this conclusion unchanged.

\begin{figure}[t]
    \centering
    \includegraphics[width=0.95\columnwidth]{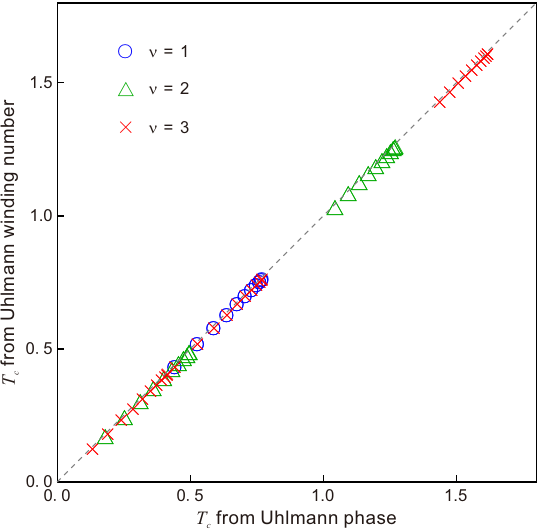}
    \caption{\textbf{Critical-temperature comparison for generalized SSH models.} Critical temperatures obtained independently from the Uhlmann phase and the Uhlmann winding number for $\nu=1,2,3$. The circles, triangles, and crosses denote the three winding numbers, respectively. For each $\nu$, the ratio $\lambda=t_0/t_\nu$ is sampled uniformly from $-0.85$ to $0.85$, with $\lambda=0$ providing the flat-band reference. Each symbol represents one thermal transition, and the dashed line indicates equal critical temperatures from the two calculations.}
    \label{Compare}
\end{figure}

\suppnote{Quasienergy spectra and Bloch vectors of the quantum walks}
\label{sec:qwspec}

Here, we provide the quasienergy spectrum and unit Bloch vector of the quantum-walk operator defined in Eq.\,(10) of the main text. 
In momentum space, the $m$-site conditional shifts are:
\begin{equation}\label{eq:qwshift}
S_{+,k}^{(m)}
=
\operatorname{diag}(e^{-imk},1),
\qquad
S_{-,k}^{(m)}
=
\operatorname{diag}(1,e^{imk}).
\end{equation}
The walk operator takes the standard SU(2) form:
\begin{equation}\label{eq:qwdecomp}
\mathcal U_{m,k}
=
\cos E_{m,k}\,\mathbb I
-i\sin E_{m,k}\,
\hat{\mathbf n}_{m,k}\cdot\boldsymbol{\sigma}.
\end{equation}
Direct multiplication of the coin and conditional-shift operators gives:
\begin{equation}\label{eq:qwspec}
\begin{aligned}
E_{m,k}
&=
\arccos\,[
-\cos\tfrac{\theta_1}{2}\cos\tfrac{\theta_2}{2}-\sin\tfrac{\theta_1}{2}\sin\tfrac{\theta_2}{2}
\cos(mk)
],
\\
\hat n^x_{m,k}
&=
[\cos\tfrac{\theta_1}{2}
\sin\tfrac{\theta_2}{2}
\sin(mk)]/\sin E_{m,k}
,
\\
\hat n^y_{m,k}
&=
[
\sin\tfrac{\theta_1}{2}
\cos\tfrac{\theta_2}{2}-\cos\tfrac{\theta_1}{2}
\sin\tfrac{\theta_2}{2}
\cos(mk)
]/\sin E_{m,k},
\\
\hat n^z_{m,k}
&=
-
[\sin\tfrac{\theta_1}{2}
\sin\tfrac{\theta_2}{2}
\sin(mk)]/\sin E_{m,k}
.
\end{aligned}
\end{equation}
The two quasienergy bands are $\pm E_{m,k}$, with $E_{m,k}\in[0,\pi]$. The expressions for $\hat{\mathbf n}_{m,k}$ are well defined away from quasienergy-gap closings, where $\sin E_{m,k}\neq0$. 
The unit Bloch vector lies in a momentum-independent plane. The unit vector normal to this plane is:
\begin{equation}\label{eq:qwplane}
\mathbf e^\perp
=
\left(
\sin\tfrac{\theta_1}{2},
0,
\cos\tfrac{\theta_1}{2}
\right),
\end{equation}
which satisfies $\hat{\mathbf n}_{m,k}\cdot\mathbf e^\perp=0$. 
The corresponding chiral operator is $\Gamma=\mathbf e^\perp\cdot\boldsymbol{\sigma}$. Because the momentum dependence enters through $mk$, a nontrivial Bloch trajectory is traversed $m$ times as $k$ spans the Brillouin zone, giving $|\nu|=m$.

\suppnote{Preparation of band eigenstates and thermal equilibrium states}
\label{sec:prep}

At the working point $(\theta_1,\theta_2)=(0,\pi)$ for $m=1,2,$ and $3$, the quasienergy spectra are flat, with $E_{m,k}=\pi/2$. The corresponding unit Bloch vectors are:
\begin{equation}\label{app:target}
\hat{\mathbf n}_{m,k}
=
\bigl(
\sin(mk),
-\cos(mk),
0
\bigr).
\end{equation}
The azimuthal angle of $\hat{\mathbf n}_{m,k}$ is $mk-\pi/2$. Consequently, the unit Bloch vector winds $m$ times as $k$ traverses the Brillouin zone.
The auxiliary walk used for state preparation is
$\mathcal U_m(0,\frac{\pi}{2})$. After omitting a scalar phase that does not affect the momentum-resolved coin projectors, its SU(2) part corresponds to a $-\frac{\pi}{2}$ rotation of the coin Bloch vector. 
The corresponding rotation axis is:
\begin{equation}\label{app:aux}
\overline{\hat{\mathbf n}}_{m,k}
=
\bigl(
\sin(mk),
-\cos(mk),
0
\bigr),
\end{equation}
which is identical to the target unit Bloch vector $\hat{\mathbf n}_{m,k}$.

The localized input states for preparing the ground-band and excited-band eigenstates can be written in momentum space as:
\begin{align}\label{app:input}
\ket{x=0,\uparrow}
=
\int_{-\pi}^{\pi}
\frac{dk}{\sqrt{2\pi}}\,
\ket{k}\otimes\ket{\uparrow},\\
\ket{x=0,\downarrow}
=
\int_{-\pi}^{\pi}
\frac{dk}{\sqrt{2\pi}}\,
\ket{k}\otimes\ket{\downarrow},
\end{align}
respectively. Hence, the same input coin state is applied uniformly to every momentum component.


Let $\mathbf s_{g,m,k}$ and $\mathbf s_{e,m,k}$ be the corresponding output Bloch vectors of the auxiliary walk.
For the target Hamiltonian
$H_{m,k}=E_{m,k}\hat{\mathbf n}_{m,k}\cdot\boldsymbol{\sigma}$, the ground- and excited-band Bloch vectors are $\mathbf n_{g,m,k}=-\hat{\mathbf n}_{m,k}$ and $\mathbf n_{e,m,k}=\hat{\mathbf n}_{m,k}$, respectively. The angular difference between $\mathbf s_{g,m,k}$ ($\mathbf s_{e,m,k}$) and $\mathbf n_{g,m,k}$ ($\mathbf n_{e,m,k}$) is $\pi/2$, which is independent of momentum and $m$. 
This difference can be compensated by a $+\pi/2$ rotation about the $z$ axis.
These momentum-independent rotations about the $z$ axis are implemented by the programmable phase shifters.
After the phase compensation, the inputs $\ket{x=0,\uparrow}$ and $\ket{x=0,\downarrow}$ generate the ground- and excited-band projectors, respectively, at each momentum.

To prepare the band eigenstates of the extended Hamiltonian $H_{m,k}(\vartheta)$, we replace the auxiliary walk operator by $\mathcal U_m(0,\pi-\vartheta)$ while keeping the localized input states unchanged. This operation rotates the coin Bloch vectors by $\vartheta-\pi$ about the axis $\overline{\hat{\mathbf n}}_{m,k}$. 
After the same $+\pi/2$ rotation about the $z$ axis, the two outputs correspond to the ground- and excited-band eigenstate textures of $H_{m,k}(\vartheta)$, respectively.

For the original 1D Hamiltonian at $\vartheta=\pi/2$, incoherently combining the ground- and excited-band outputs with $p_g(T)=[1+\tanh(E/T)]/2$ and $p_e(T)=[1-\tanh(E/T)]/2$ yields the target thermal state $\rho_k(T)$.
Because the dimensional extension preserves the flat quasienergy $E=\pi/2$, the same Boltzmann weights remain independent of both momentum and $\vartheta$.
Applying the same incoherent mixing procedure to the extended band eigenstates therefore yields $\rho_k(\vartheta,T)$, with $\rho_k(\pi/2,T)=\rho_k(T)$.


\suppnote{From real-space measurements to momentum-space topological invariants}
\label{sec:recon}

In this section, we describe how the position--coin density matrices reconstructed by quantum-state tomography are transformed into momentum-resolved density matrices. 
We then detail the numerical Uhlmann parallel transport used to evaluate the full-loop and momentum-resolved Uhlmann phases, as well as the winding of the extended-overlap curves.

\newcounter{suppsubsec}[suppnote]
\newcommand{\suppsubsection}[1]{%
  \refstepcounter{suppsubsec}%
  \subsection*{\arabic{suppnote}.\arabic{suppsubsec}\quad #1}%
}

\suppsubsection{Momentum-space reconstruction}

Quantum-state tomography yields the position--coin density matrix $\rho_{\mathrm{exp}}(T)$. Its position-space blocks are defined as $\rho_{x,x'}(T)=\langle x|\rho_{\mathrm{exp}}(T)|x'\rangle$, each of which is a $2\times2$ matrix in the coin space. The diagonal momentum-space block is obtained by Fourier transformation:
\begin{equation}\label{eq:momentum_reconstruction}
\widetilde{\rho}_k(T)
=
\sum_{x,x'}
e^{+ik(x-x')}
\rho_{x,x'}(T).
\end{equation}
The normalized momentum-resolved density matrix is:
\begin{equation}\label{eq:momentum_normalization}
\rho_k(T)
=
\frac{\widetilde{\rho}_k(T)}
{\operatorname{Tr}[\widetilde{\rho}_k(T)]}.
\end{equation}
This procedure extracts the $k=k'$ block of the Fourier-transformed position--coin density matrix.

\suppsubsection{Numerical evaluation of the Uhlmann phase}
\label{sec:Ucalc}

For each temperature $T$, the reconstructed momentum loop is sampled at $N=100$ equally spaced points $k_i\in[-\pi,\pi)$, with $k_{N+1}=k_1$. For compactness, we write $\rho_i=\rho_{k_i}(T)$. The positive square root of each density matrix is obtained from its spectral decomposition:
\begin{align}
\rho_i
&=
\sum_{\alpha=\pm}
\lambda_{\alpha,i}
\ketbra{u_{\alpha,i}}{u_{\alpha,i}},
\nonumber\\
\sqrt{\rho_i}
&=
\sum_{\alpha=\pm}
\sqrt{\lambda_{\alpha,i}}
\ketbra{u_{\alpha,i}}{u_{\alpha,i}}.
\end{align}

For each neighboring pair, including the link that closes the momentum loop, we perform the polar decomposition:
\begin{align}\label{eq:polar}
\sqrt{\rho_{i+1}}\sqrt{\rho_i}
&=
V_iP_i,
\nonumber\\
P_i
&=
\left(
\sqrt{\rho_i}\rho_{i+1}\sqrt{\rho_i}
\right)^{1/2},
\end{align}
where $V_i$ is the unitary polar factor and $P_i$ is positive semidefinite. Writing the Uhlmann amplitudes as $w_i=\sqrt{\rho_i}U_i$, we choose $U_1=\mathbb I$ and propagate the gauge according to $U_{i+1}=V_iU_i$. This gives:
\begin{equation}\label{eq:parallel}
w_{i+1}^{\dagger}w_i
=
U_i^{\dagger}P_iU_i
\geq 0,
\end{equation}
which is the discrete Uhlmann parallel-transport condition.

After completing the closed momentum loop, the Uhlmann holonomy is:
\begin{equation}\label{eq:discrete_holonomy}
U_{\mathrm{Uhl}}(T)
=
V_N(T)V_{N-1}(T)\cdots V_1(T).
\end{equation}
The full-loop Uhlmann overlap and phase are then:
\begin{equation}\label{eq:discrete_phase}
\mathcal G_U(T)
=
\operatorname{Tr}
\left[
\rho_{k_1}(T)U_{\mathrm{Uhl}}(T)
\right],
\qquad
\Phi_U(T)=\arg\mathcal G_U(T).
\end{equation}

To obtain the momentum-resolved phase, we use the partial holonomy $U_j=V_{j-1}\cdots V_1$ and evaluate:
\begin{align}\label{eq:partial_overlap}
\mathcal G_U(k_j,T)
&=
\operatorname{Tr}
\left[
\sqrt{\rho_1}\sqrt{\rho_j}U_j
\right],
\nonumber\\
\Phi_U(k_j,T)
&=
\arg\mathcal G_U(k_j,T).
\end{align}
Upon completion of the loop, these quantities reduce to the full-loop overlap $\mathcal G_U(T)$ and Uhlmann phase $\Phi_U(T)$.

\suppsubsection{Numerical evaluation of the Uhlmann winding number}
\label{sec:UWcalc}

For each temperature $T$, the auxiliary interval is sampled at $\vartheta_j=j\pi/(N_\vartheta-1)$, with $j=0,\ldots,N_\vartheta-1$, where $N_\vartheta$ denotes the number of sampled points.
The two endpoints satisfy $\vartheta_0=0$, $\vartheta_{N_\vartheta-1}=\pi$, and $\widetilde{\mathcal G}_U(\vartheta_0,T)=\widetilde{\mathcal G}_U(\vartheta_{N_\vartheta-1},T)=1$.
Away from the critical temperatures, the discrete Uhlmann winding number is evaluated as:
\begin{equation}\label{eq:discrete_UWN}
\nu_U^{\mathrm{disc}}(T)
=
\frac{1}{2\pi}
\sum_{j=0}^{N_\vartheta-2}
\operatorname{Arg}\!\left[
\frac{
\widetilde{\mathcal G}_U(\vartheta_{j+1},T)
}{
\widetilde{\mathcal G}_U(\vartheta_j,T)
}
\right].
\end{equation}
Here, $\operatorname{Arg}$ takes values in $(-\pi,\pi]$, and each term gives the oriented phase difference between two neighboring overlap points.
The sampling is chosen sufficiently dense that the continuous phase change between neighboring points remains smaller than $\pi$.
Eq.\,\eqref{eq:discrete_UWN} then gives the winding of the discretized closed overlap trajectory around the zero-overlap point.

We set $\Delta\widetilde{\Phi}_U(\vartheta_0,T)=0$. For $j=1,\ldots,N_\vartheta-1$, the accumulated continuously unwrapped phase is evaluated as:
\begin{equation}\label{eq:discrete_unwrapped_phase}
\Delta\widetilde{\Phi}_U(\vartheta_j,T)
=
\sum_{\ell=0}^{j-1}
\operatorname{Arg}\!\left[
\frac{
\widetilde{\mathcal G}_U(\vartheta_{\ell+1},T)
}{
\widetilde{\mathcal G}_U(\vartheta_\ell,T)
}
\right].
\end{equation}
Its endpoint value satisfies $\Delta\widetilde{\Phi}_U(\vartheta_{N_\vartheta-1},T)=2\pi\nu_U^{\mathrm{disc}}(T)$.
At a critical temperature, the overlap trajectory passes through the zero-overlap point, where the overlap phase and winding number become ill-defined.

\suppnote{Calibration details and characterization results}
\label{app:chip_cal}
\begin{figure}
    \centering
    \includegraphics[width=0.9\linewidth]{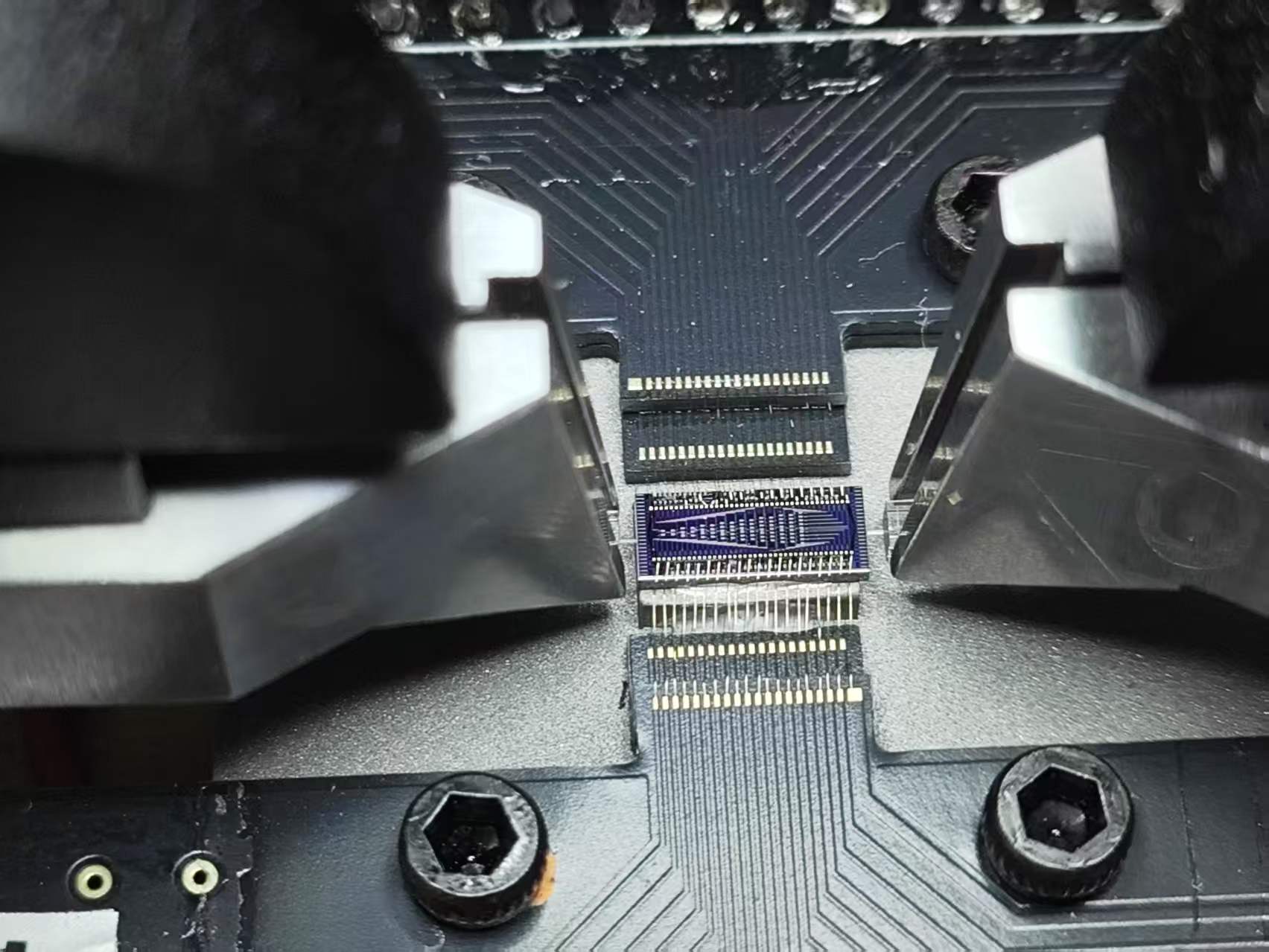}
    \caption{\textbf{Photonic-chip packaging and fiber coupling.} The silicon photonic chip is wire-bonded to a printed circuit board and coupled to the input and output fiber arrays through edge couplers.}
    \label{fig:coupling}
\end{figure}

\begin{figure*}
    \centering
    \includegraphics[width=1\linewidth]{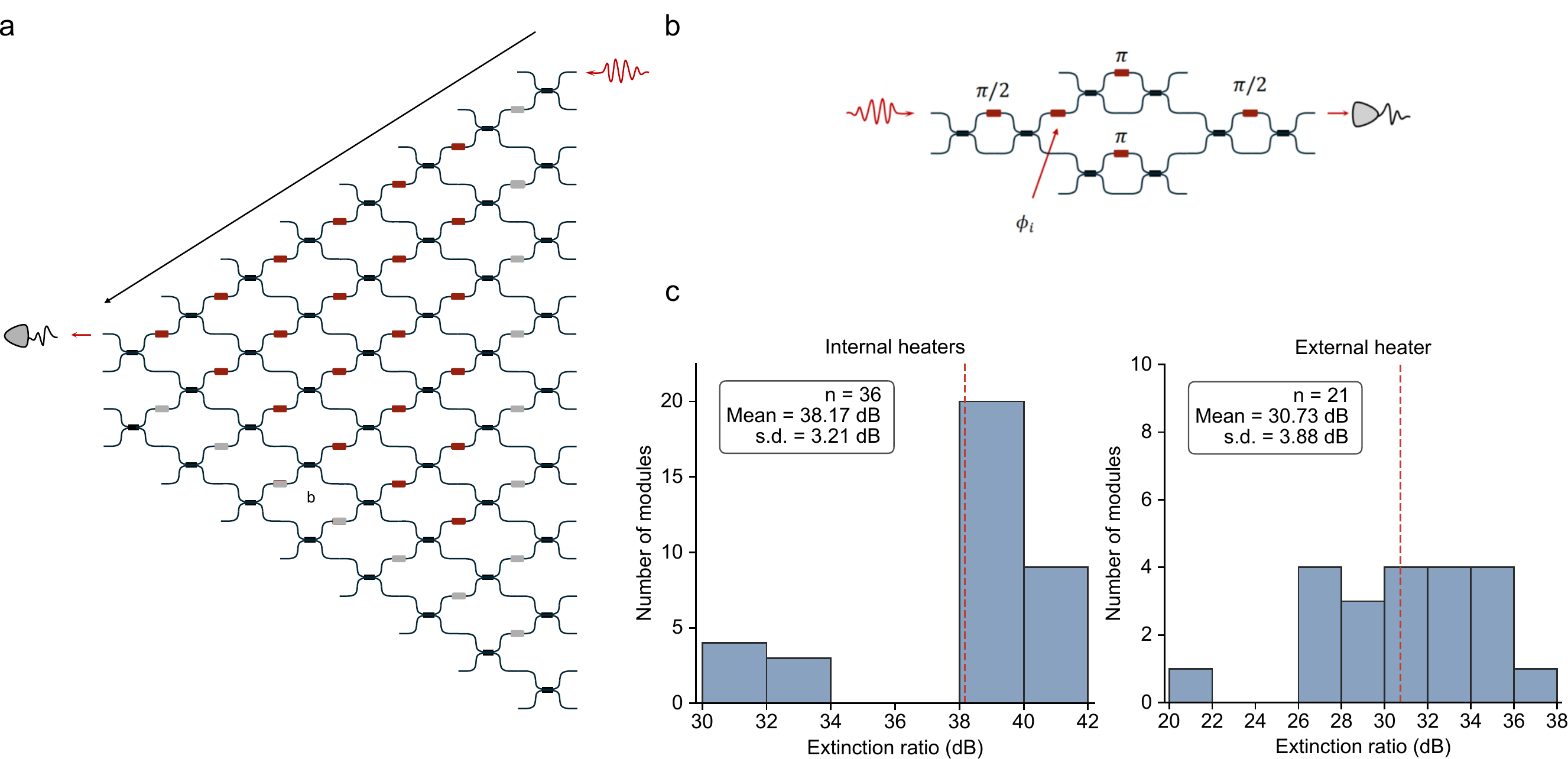}
    \caption{
\textbf{Calibration and characterization of the thermo-optic phase shifters.}
\textbf{a}, Diagonal-path calibration of the internal phase shifters. Each black mixing element represents a programmable MZI, with its internal phase shifter omitted for clarity. Red and gray rectangles denote external phase shifters used and unused in the experiments, respectively. Previously calibrated MZIs along the selected path are held at their maximum-transmission settings while the target internal phase shifter is scanned.
\textbf{b}, Effective interferometer used to calibrate an external phase shifter $\phi_i$. The internal phases of the outer MZIs are set to $\pi/2$ for balanced splitting and recombination, while those of the intermediate MZIs are set to $\pi$ to route light through the interferometer arms.
\textbf{c}, Extinction-ratio distributions for 36 internal-phase-shifter MZIs and 21 external-phase-shifter test structures. Red dashed lines indicate the corresponding mean values.
}
    \label{fig:heater_ER}
\end{figure*}

As shown in Supplementary Fig.\,\ref{fig:coupling}, the chip was mounted and wire-bonded to a printed circuit board for electrical control. Single-mode fiber arrays were aligned with the edge couplers for optical input and output.

For the internal phase shifters, a diagonal optical path through the mesh was selected using an appropriate pair of calibration input and output ports, as illustrated in Supplementary Fig.\,\ref{fig:heater_ER}(a). At each calibration step, previously calibrated Mach--Zehnder interferometer (MZIs) along the selected path were held at settings that maximized transmission along that path. The current applied to the target heater was then swept, and the corresponding output power was recorded. The resulting interference fringe yielded the phase--current response, with the maximum-transmission point taken as the phase reference. The calibrated MZI was subsequently fixed at this setting before proceeding to the next unit. This procedure was repeated over successive diagonals until all internal phase shifters used in the experiments had been calibrated.

Supplementary Fig.~\ref{fig:heater_ER}(b) shows the effective interferometer used to calibrate an external phase shifter $\phi_i$. The internal phases of the two outer MZIs were set to $\pi/2$ to provide balanced splitting and recombination, while those of the two intermediate MZIs were set to $\pi$ to route light through the two interferometer arms. These four MZIs were held at their prescribed settings while the current applied to the target external heater was swept. The resulting output interference fringe was used to determine the phase--current response of the external phase shifter.

For both calibration configurations, the extinction ratio was determined as
\begin{equation}
    \mathrm{ER}
    = 10\log_{10}\left(\frac{P_{\max}}{P_{\min}}\right),
\end{equation}
where $P_{\max}$ and $P_{\min}$ denote the maximum and minimum output powers of the interference fringe, respectively. For an internal phase shifter, the extinction ratio characterizes the corresponding MZI; for an external phase shifter, it characterizes the effective interferometer formed by the surrounding MZIs.

The measured extinction-ratio distributions are shown in Supplementary Fig.\,\ref{fig:heater_ER}(c). They comprise 36 internal-phase-shifter MZIs and 21 external-phase-shifter test structures, covering all 57 phase shifters used in the experiments. The total of 79 reported in the Methods section of the main text includes all phase shifters fabricated on the chip. In the simplified mesh schematic in Supplementary Fig.\,\ref{fig:heater_ER}(a), the internal phase shifters are contained within the MZI units and are not shown individually; red and gray rectangles denote external phase shifters used and unused in the experiments, respectively.


\end{document}